\documentclass{amsart}
\usepackage{epsfig}
\usepackage{amsmath,amssymb}
\usepackage{graphics}
\newtheorem{theorem}{Theorem}

\newtheorem{proposition}[theorem]{Proposition}

\begin{document}
\title[Existence of non-abelian monopoles]{On the existence of non-abelian monopoles: the algebro-geometric approach}
\author{H.W. Braden}
\address{School of Mathematics, Edinburgh University, Edinburgh.}
\email{hwb@ed.ac.uk}
\author{V.Z. Enolski}
\address{ZARM, Universit\"at Bremen, Am Fallturm, D-28359, Bremen\\
On leave from: Institute of Magnetism, National Academy of Sciences of
Ukraine.} \email{vze@ma.hw.ac.uk}

\begin{abstract}
We develop the Atiyah-Drinfeld-Manin-Hitchin-Nahm construction to
study  $SU(2)$ non-abelian charge 3 monopoles within the
algebro-geometric method. The method starts with finding an
algebraic curve, the monopole spectral curve,  subject to
Hitchin's constraints. We take as the monopole curve the genus
four curve that admits a $C_3$ symmetry,
$\eta^3+\alpha\eta\zeta^2+\beta\zeta^6+\gamma\zeta^3-\beta=0$,
with real parameters $\alpha$, $\beta$ and $\gamma$. In the case
$\alpha=0$ we prove that the only suitable values of
$\gamma/\beta$ are $\pm 5\sqrt{2}$ ($\beta$ is given below) which
corresponds to the tetrahedrally symmetric solution. We then
extend this result by continuity to non-zero values of the
parameter $\alpha$ and find finally a {\em new} one-parameter
family of monopole curves with $C_3$ symmetry.
\end{abstract}

\maketitle

{\bf 1. What is the monopole?} Non-abelian monopoles appear
naturally as particular solutions within the Standard Model, see
e.g. the recent review \cite{weinbyi06} and the monograph
\cite{ms04}. The associated Lagrangian density in Minkowski space
is given by
\begin{align} L=-\frac14\mathrm{Tr}\, F_{ij} F^{ij}+  \frac12\mathrm{Tr}\,D_{i}\Phi D^{i}\Phi + V.
\label{lagrangian}\end{align} Here $F_{ij}$ is the Yang-Mills
field strength,
\begin{align}
F_{ij}=\partial_i a_j -\partial_j a_i+[a_i,a_j],\label{Yang-Mills}
\end{align}
$a_j$ the corresponding gauge field, $D_i$ the associated
covariant derivative acting on the Higgs field $\Phi$ by
$$D_i\Phi=\partial_i\Phi+[a_i,\Phi],$$
and $V$ a potential. The gauge and Higgs fields take values in the
Lie algebra of the gauge group. Static finite energy solutions of
the Model are supposed to satisfy to the boundary conditions
\begin{align}&\left.\sqrt{-\frac12\mathrm{Tr}\; \Phi(r)^2}
\right|_{r\rightarrow\infty}\sim 1-\frac{n}{2r}+O(r^{-2}), \quad \text{with} \quad r=\sqrt{x_1^2+x_2^2+x_3^2}.\label{charge}
\end{align}
The positive integer $n\in \mathbb{N}$ in (\ref{charge}) is the
first Chern number or the charge. Such a solution is called a {\em
non-abelian monopole of charge $n$}.

We consider here non-abelian monopoles in the BPS
(Bogomolny-Prasad-Sommerfeld)  limit for which the potential $V=0$
but the boundary conditions (\ref{charge}) remain preserved. Thus
the configurations that minimize the energy of the system solve
the {\em Bogomolny equations}
\begin{equation} D_i\Phi=\pm\sum_{j,k=1}^3\epsilon_{ijk} F_{jk},\quad i=1,2,3. \end{equation}
Moreover we fix the gauge group to be $SU(2)$: therefore our
development deals with the {\em static $SU(2)$ monopoles in the
BPS limit}. \vskip0.3cm

{\bf 2. The Atiyah-Drinfeld-Manin-Hitchin-Nahm construction.}
Although the Bogomolny equation is a first order partial
differential equation in $\mathbb{R}^3$ few explicit solutions are
known for $n>1$. Our results are based on the ADMHN construction
that reduces this partial differential equation to a completely
integrable ordinary differential equation. We summarize the
construction in the form of the following theorem.

\begin{theorem} [\bf{ADMHN}]
The $su(2)$ charge $n$ monopole solution is given by
\begin{align*}\begin{split}
\Phi(\boldsymbol{x})_{\mu\nu}&=\imath \int_{0}^2 s
\boldsymbol{v}_{\mu}^{\dagger}(\boldsymbol{x},s)
\boldsymbol{v}_{\nu}(\boldsymbol{x},s)\mathrm{d}s,\\
a_i(\boldsymbol{x})_{\mu\nu}&=\imath \int_{0}^2
\boldsymbol{v}_{\mu}^{\dagger}(\boldsymbol{x},s)
\frac{\partial}{\partial x_i}
\boldsymbol{v}_{\nu}(\boldsymbol{x},s)\mathrm{d}s,\quad
i=1,2,3,\end{split}\mu,\nu=1,2.
\end{align*}
where $\boldsymbol{v}_{\mu}(\boldsymbol{x},s)$ are two
orthonormalizable
\[\int_{0}^2 \boldsymbol{v}_{\mu}^{\dagger}(\boldsymbol{x},s)
\boldsymbol{v}_{\nu}(\boldsymbol{x},s)\mathrm{d}s=\delta_{\mu\nu}
,\quad \mu,\nu=1,2,\]
solutions to the Weyl equation
\begin{align}\left(-\imath 1_{2n}\frac{\mathrm{d}}{\mathrm{d}s} +\sum_{j=1}^3( T_j(s)+\imath x_j 1_n) \otimes
\sigma_j \right)
\boldsymbol{v}(\boldsymbol{x},s)=0. \label{Weyl}
\end{align}
The $n\times n$ matrices $T_j(s)$, $s\in (0,2)$, called Nahm data,
satisfy the Nahm equations
\begin{equation} \frac{\mathrm{d}T_i(s)}{\mathrm{d}s}
=\frac12\sum_{j,k=1}^3\epsilon_{ijk}[T_j(s),T_k(s)].\label{Nahm}
\end{equation}
The residues $\mathrm{Res}_{s=0}T_i(s)$ and
$\mathrm{Res}_{s=2}T_i(s)$ form irreducible $n$-dimensional
repre\-sen\-tations of $su(2)$. Also the following hermiticity
conditions are satisfied
$$T_i(s)=-T_i^{\dagger}(s), \qquad T_i(s)=T_i^{\dagger}(2-s).$$
\end{theorem}
\vskip0.3cm {\bf 3. The Hitchin construction.} The complete
integrability of Nahm's equations (\ref{Nahm}) was proved by
Hitchin in \cite{hitchin82}. These equations can be written the
Lax form,
\begin{align}\frac{\mathrm{d}A(s,\zeta)}
{\mathrm{d}s}=[A(s,\zeta),M(s,\zeta)],\label{Lax}\end{align}
where $\zeta$ is a spectral parameter and $A(s,\zeta)$, $M(s,\zeta)$ are $n\times n$ matrices
\begin{align*}
 &A(s,\zeta)=A_{-1}(s)\zeta^{-1}+A_0(s)+A_{+1}(s)\zeta,\\ &M(s,\zeta)=\frac12 A_0(s)+\zeta A_{+1}(s),\\
 &A_{\pm1}(s)=T_1(s)\pm\imath T_2(s),\quad A_0(s)=2\imath T_3(s). \end{align*}
The known consequence of the Lax representation is that the equation
\begin{equation}  \mathrm{det} (A(s,\zeta)-\eta 1_n) =0 \label{determinant}\end{equation}
represents a polynomial in $(\eta,\zeta)$ with coefficients
independent of $s$. That is the spectral curve
$\widehat{\mathcal{C}}$ is given by the equation
 \begin{equation}\eta^n+\alpha_1(\zeta)\eta^{n-1}+\ldots+\alpha_n(\zeta)=0, \label{curve}\end{equation}
where $\alpha_k(\zeta)$ are polynomials in $\zeta$ of degree not
exceeding $2k$. The genus of $\widehat{\mathcal{C}}$ is
generically
 \begin{equation}{g}_{\widehat{\mathcal{C}}}=(n-1)^2.\label{genus} \end{equation}

\vskip0.3cm {\bf 4. Theta-functions.} Riemann's $\theta$-function
is a powerful instrument of analysis of algebraic curves and their
Jacobians, see e.g. \cite{fay73}.  $\theta$-functions depend on
two groups of variables, a complex vector $\boldsymbol{z}\in
\mathbb{C}^g$ and a period matrix belonging to the Siegel upper
half-space
\begin{equation}\widehat{\tau}:\quad  \widehat{ \tau}^T=\widehat{\tau},\quad \Im(\widehat{\tau})>0. \label{tau}
\end{equation}
The period matrix $\widehat{\tau}$ is built from a complete set of
linearly independent holomorphic differentials
\begin{equation}\boldsymbol{u}(\xi,\eta)=(u_1(\xi,\eta),\ldots, u_g(\xi,\eta))^T\label{differentials}\end{equation}
and a canonical homology basis
\begin{equation}(\mathfrak{a}_1,\ldots,\mathfrak{a}_g;
\mathfrak{b}_1,\ldots,\mathfrak{b}_g), \quad \mathfrak{a_i}\circ
\mathfrak{b_j}=\delta_{i,j},\; \mathfrak{a_i}\circ \mathfrak{a_j}=
 \mathfrak{b_i}\circ \mathfrak{b_j}=0.\end{equation}
Denoting the matrices of $\mathfrak{a}$ and $\mathfrak{b}$-periods
as
\begin{align*}
\mathcal{A}&= \left(\oint_{\mathfrak{a}_i} u_j(\xi,\eta)\right)_{i,j=1,\ldots,g} ,\;
 \mathcal{B}= \left(\oint_{\mathfrak{b}_i} u_j(\xi,\eta)\right)_{i,j=1,\ldots,g}.
\end{align*}
we then define $\widehat{\tau}=  \mathcal{B}\mathcal{A}^{-1}$. The
$\theta$-function of the algebraic curve $\widehat{\mathcal{C}}$
is given by the Fourier series
\begin{equation} \theta(\boldsymbol{z};\widehat{\tau})=\sum_{\boldsymbol{n}\in \mathbb{Z}^g}
\mathrm{exp}\left\{ \imath\pi \boldsymbol{n}^T \widehat{\tau}
\boldsymbol{n}+2\imath \pi \boldsymbol{z}^T \boldsymbol{n}
\right\}.\label{theta} \end{equation} $\theta$-functions possesses
periodicity properties when the argument is shifted by a period,
and modular properties when the homology basis is mapped  to
another one. We do not present these well-known formulae here.

\vskip0.3cm {\bf  5. Hitchin's constraints.} Not all curves of the
form (\ref{curve}) can serve as a spectral curve of a monopole but
only those that satisfy the {\em Hitchin constraints}, denoted
below as $\bf H1$, $\bf H2$ and $\bf H3$. These constraints were
formulated in \cite{hitchin82,hitchin83} as conditions on the
cohomology groups of holomorphic line bundles associated to the
spectral curve. Here we will present these conditions in
equivalent form by following to the Ercolani-Sinha paper
\cite{ersi89} and our preprint \cite{bren06} that is published in
journal form in \cite{bren10a} and \cite{bren10b}.

{\bf H1.} The spectral curve $\hat{\mathcal{C}}$ admits the involution \begin{equation}
(\zeta,\eta)\rightarrow \left( -{1}/{\overline{\zeta}}, -
{\overline{\eta}}/{\overline{\zeta}^2}\right). \label{involution}\end{equation}

{\bf H2.} The $\mathfrak{b}$-periods of a normalized differential
of the second kind $\gamma_{\infty}(P)$ are half-integer, where
\begin{align}
&\left.\gamma_{\infty}(P)\right._{P\to\infty_i}
=\left(\frac{\rho_i}{\xi^2}+O(1)\right)\mathrm{d}\xi, \quad \rho_i=\lim_{\zeta\to \infty_i} \frac{\eta}{\zeta^2},
\\& \oint_{\mathfrak{a}_k}\gamma_{\infty}=0,\quad k=1,\ldots,g,\\
 &\boldsymbol{U}=\frac{1}{2\pi\imath}
\left(\oint_{\mathfrak{b}_1}\gamma_{\infty},\ldots,
\oint_{\mathfrak{b}_g}\gamma_{\infty}  \right)^T =\frac12
\boldsymbol{n}+\frac12 \tau \boldsymbol{m}.
\end{align}
Here the integer vectors $\boldsymbol{n},\boldsymbol{m}\in
\mathbb{Z}^g$ are the {\em Ercolani-Sinha vectors} that were
introduced in \cite{ersi89} and will play the role of the
principal variables in this exposition.

As noted in \cite{ersi89}, the constraint $\bf H2$ is a very
restrictive condition on the moduli of the curve and \emph{a
priori} it is not clear if such a curve exists for $n>2$. It
places ${g}_{\widehat{\mathcal{C}}}$ real constraints on the
coefficients of (\ref{curve}).

{\bf H3.} The linear winding vector
$\boldsymbol{U}s+\boldsymbol{K}$, where $\boldsymbol{K}$ is the
vector of Riemann constants, does not intersect theta-divisor
inside the interval $(0,2)$:
\begin{equation}\quad \theta( \boldsymbol{U}s+\boldsymbol{K};\widehat{\tau} )
 \neq 0, \quad s\in (0,2).
\label{divisor}\end{equation}

\vskip0.3cm

{\bf 6. Existence of the tetrahedral monopole.} We will restrict
our analysis to the special class of curves  that respect the
$C_3$ symmetry,
\begin{equation}  \sigma:(\eta,\zeta)\longrightarrow (\rho\eta,\rho\zeta),
\quad \rho=\mathrm{e}^{2\imath\pi/3}. \label{C3} \end{equation}
This symmetry corresponds to a space-time symmetry of the monopole
\cite{hmm95}. The most general charge 3 monopole curve that admits
such a $C_3$ symmetry and satisfies $\bf H1$ may be put in the
form
\begin{align}
\eta^3+\alpha\eta\zeta^2+\beta\zeta^6+\gamma\zeta^3-\beta=0, \label{C3curve}
\end{align}
where $\alpha,\beta,\gamma$ are real numbers. We start by
considering an even more special subclass of $C_3$ symmetric
curves, namely
\begin{equation} \eta^3+\chi(\zeta^6+b\zeta^3-1)=0 . \label{bcurve}\end{equation}
We report here the following
\begin{theorem}[\bf On the existence of tetrahedral monopole \cite{bren10}]
The class of the monopole curves (\ref{bcurve}) contains the two
representatives,
\begin{equation}b=\pm 5\sqrt{2}, \qquad \chi=-\frac16\frac{\Gamma(1/6)\Gamma(1/3)}{2^{1/6}\pi^{1/2}}.\label{terahedron}\end{equation}
\end{theorem}

\vskip0.3cm {\bf 7. Demonstration.} Our proof is based on various
results, old and new. \vskip0.3cm {\bf 7.1 Wellstein and
Matsumoto.} Consider the curve of genus $g=4$
\begin{equation} w^3=(z-\lambda_1)\ldots (z-\lambda_6),\quad \lambda_i\neq \lambda_j \in \mathbb{C} \label{welstein1}\end{equation}
The associated holomorphic differentials are
\[ \frac{\mathrm{d}z}{w},\;\;\; \frac{\mathrm{d}z}{w^2},\;\;\;
 \frac{z\mathrm{d}z}{w^2},\;\;\;\frac{z^2\mathrm{d}z}{w^2}. \]
Let
$\{\mathfrak{a}_1,\ldots,\mathfrak{a}_4;\mathfrak{b}_1,\ldots,\mathfrak{b}_4\}$
be the homology basis shown in Fig.1
\begin{center}
\begin{figure}
\begin{minipage}{160pt}%\vspace*{-2mm}\hspace*{-7mm}
 \includegraphics[width=190pt]{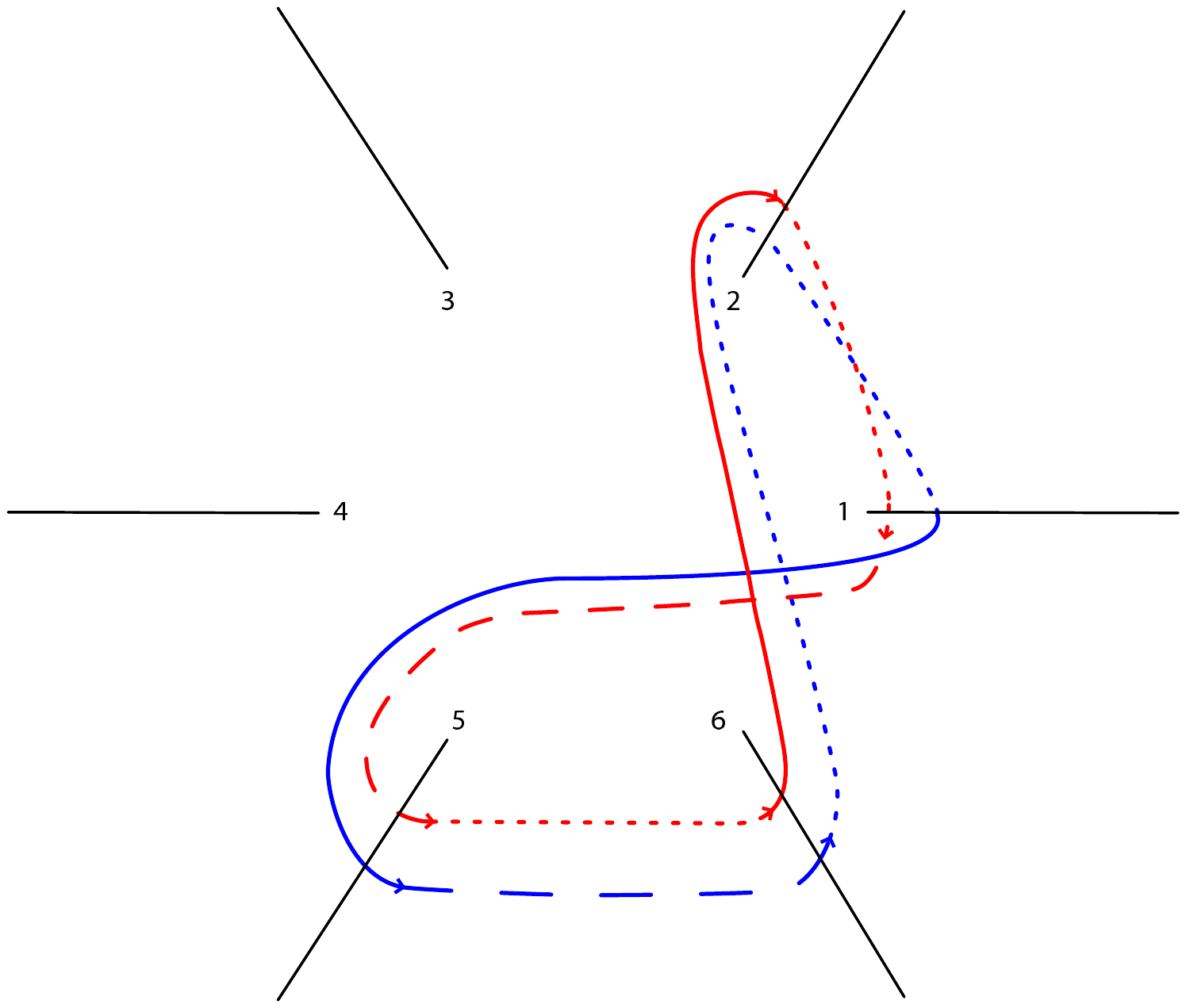}
\end{minipage}
\begin{minipage}{160pt}\vspace*{0.5in}\hspace*{-7mm}
 \includegraphics[width=190pt]{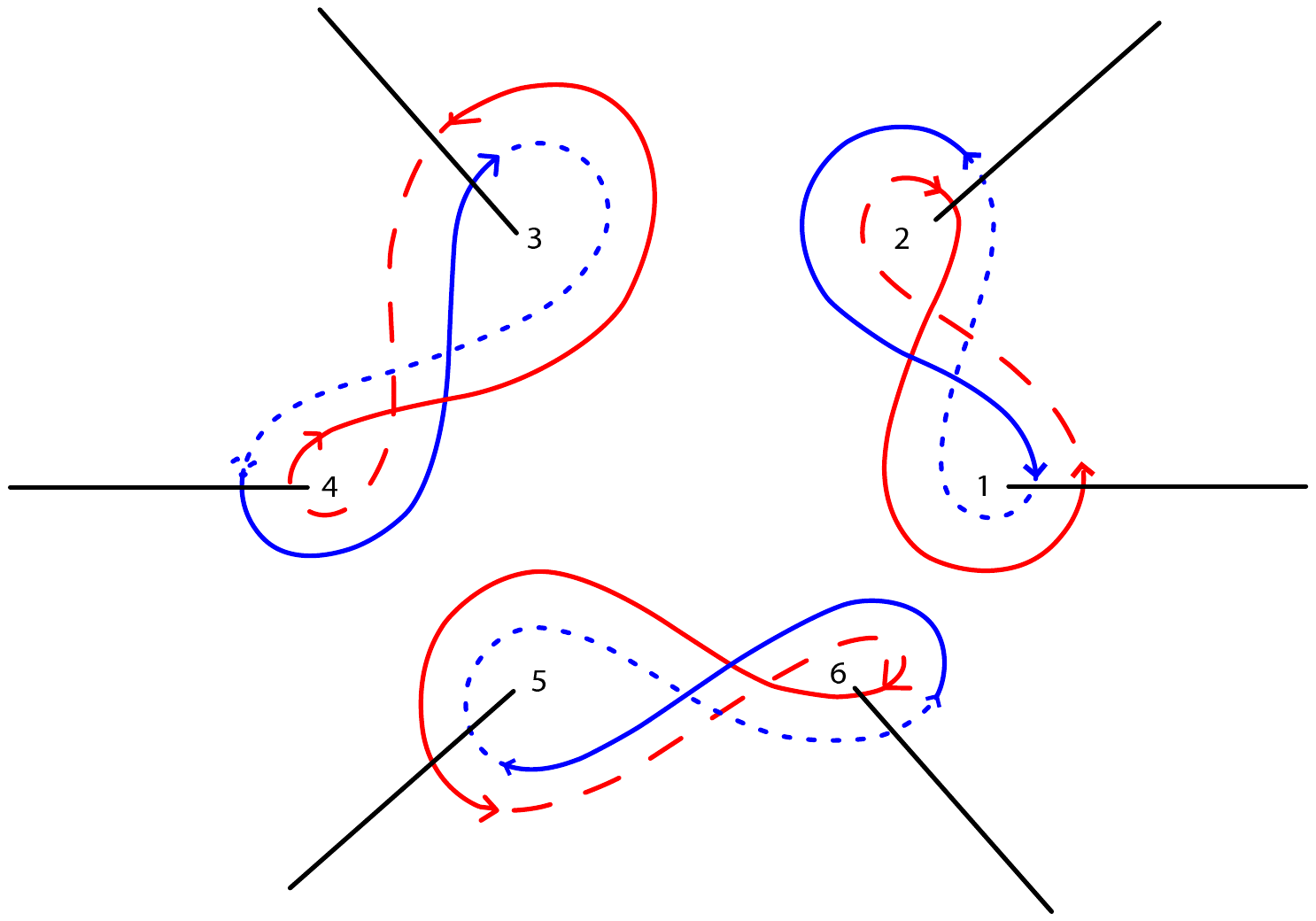}
\end{minipage}
\label{Fig.1}
\vspace*{-1in}
\caption{Homology basis of the genus four curve
$ w^3=(z-\lambda_1)\ldots (z-\lambda_6)$.
The cuts connect three sheets and are connected at infinity.
Arcs are depicted by solid lines on the first (upper) sheet,
dotted lines on the second sheet and dashed lines on the third sheet}
\end{figure}
\end{center}
Denote the vector of periods
\[ \boldsymbol{X}=\left( \oint_{\mathfrak{a}_1} \frac{\mathrm{d}z}{w},\ldots,
\oint_{\mathfrak{a}_4} \frac{\mathrm{d}z}{w} \right)^T.   \] In
1899 Wellstein showed  \cite{wel99} that the period matrix
$\widehat{\tau}$ is of the form
\begin{equation} \widehat{\tau}=\rho^2\left( H+(\rho^2-1)
 \frac{H\boldsymbol{X}\boldsymbol{X}^TH}{\boldsymbol{X}^TH\boldsymbol{X}}
 \right),\label{tau} \end{equation}
where $\rho=\mathrm{exp}(2\imath \pi/3)$,
$H=\mathrm{diag}(1,1,1,-1)$. This was rediscovered by Matsumoto in
2000 \cite{matsu00} and a further proof given in \cite{bren06}. We
will implement Wellstein's result in the case of the special curve
of the form (\ref{C3curve}) given by (\ref{bcurve}). This is still
of genus four.

It was shown in \cite{bren06} that for a pair of relatively prime
integers $(m,n)$ for which $(m+n)(m-2n)<0$ the following  solution
to {\bf H1} and {\bf H2} could be constructed. First one solves
for $t$ the equation involving hypergeometric functions
\[ \frac{2n-m}{m+n} =\frac{{}_2F_1\left(\frac13,\frac23;1,t\right)}
{{}_2F_1\left(\frac13,\frac23;1,1-t\right)}.  \] Then values of
parameters $b$ and $\chi$ are given by
\begin{align}
\begin{split}
b&=\frac{1-2t}{\sqrt{t(1-t)}},\\
\chi^{1/2}&=-(n+m)\frac{2\pi}{3\sqrt{3}}
\sqrt[3]{t^2(1-t)}\;{}_2F_1\left(\frac13,\frac23;1,t\right).\end{split}
\label{parameters}  \end{align} The Ercolani-Sinha vectors are
then expressible in terms of two integers $m,n\in\mathbb{Z}$,
\begin{equation} \boldsymbol{n}=\left(\begin{array}{c} n\\ m-n\\ -m\\2n-m
\end{array}\right), \quad
\boldsymbol{m}=\left(\begin{array}{c} -m\\ n\\ m-n\\3n
\end{array}\right). \label{esvector} \end{equation} The period
matrix then takes the form
\begin{equation} \widehat{\tau}=\rho\sp2 H+(\rho-\rho\sp2)
\frac{(\boldsymbol{n}+\rho^2H\boldsymbol{m})
(\boldsymbol{n}+\rho^2H\boldsymbol{m})^T} {(\boldsymbol{n}
+\rho^2H\boldsymbol{m})^TH
(\boldsymbol{n}+\rho^2H\boldsymbol{m})}.\end{equation} i.e. it
depends on the integers $(m,n)$ and root of unity $\rho$.

\vskip0.3cm { \bf 7.2 A Strange equation.} Comparing our
parametrization with Hitchin, Manton and Murray's  tetrahedral
solution \cite{hmm95} we conclude that with $n=1$ and $m=0$ we
should should have
\begin{align}
 \frac {{}_2F_1\left(\frac13,\frac23;1; t   \right)}
{{}_2F_1\left(\frac13,\frac23;1; 1-t  \right)}=2,\label{strange1}\\ t
=\frac12+\frac{5\sqrt{3}}{18},\quad b=-5\sqrt{2}.\label{strange2}
\end{align}
One can see that although equation (\ref{strange1}) is
transcendental it is nonetheless solved in radicals
 (\ref{strange2}). There is also a solution which corresponds to a physical inversion of
the $n=1$, $m=0$ solution with $n=m=1$: for this latter solution
$t =\frac12-\frac{5\sqrt{3}}{18}$, and $b=5\sqrt{2}$. We will
focus on the former solution in the ensuing discussion.

The following question arises: can one find numbers $t$ such that
\[ \frac {{}_2F_1\left(\frac13,\frac23;1; t  \right)}
{{}_2F_1\left(\frac13,\frac23;1; 1-t
\right)}=\frac{2n-m}{n+m}\in\mathbb{Q}?\] Each such solution will
then provide a curve satisfying $\bf H1$ and $\bf H2$. \vskip0.3cm
{\bf 7.3 Ramanujan's hypergeometric relation.} The answer to the
question just posed follows from Ramanujan's hypergeometric
relation presented in the Second Notebook \cite{berndt98}.

Let $r$ (the signature) and $n\in \mathbb{N}$. Then the following
hypergeometric equality holds when $x,y$ are the zeros of a
(necessarily) algebraic equation $\mathcal{P}(x,y)=0$,
\begin{equation}
\frac {{}_2F_1\left(\frac{1}{r},\frac{r-1}{r};1; 1-x   \right)}
{{}_2F_1\left(\frac{1}{r},\frac{r-1}{r};1; x   \right)}=n\, \frac
{{}_2F_1\left(\frac{1}{r},\frac{r-1}{r};1;1- y   \right)}
{{}_2F_1\left(\frac{1}{r},\frac{r-1}{r};1;y    \right)}
.\label{Ramanujan}\end{equation}  A consequence of this is that
the numbers $t$ above are {\em algebraic}. Ramanujan found this
equation for signature $r=3$ and $n=2$ where
\begin{equation}(xy)^{\frac13}+(1-x)^{\frac13}(1-y)^{\frac13}=1.\label{Ram3} \end{equation}
Setting $y=\frac12$ in (\ref{Ram3}) we  obtain $x=\frac12\mp
\frac{5\sqrt{3}}{18}$ and $b=\pm5\sqrt{2}$. Therefore Ramanujan's
relation stands behind the existence of the tetrahedral monopole!

To complete the proof of the existence of a monopole spectral
curve it remains to check that the curves satisfying $\bf H1$ and
$\bf H2$ also satisfy $\bf H3$, i.e. to show that for $s\in(0,2)$
the winding vector does not intersect the $\theta$-divisor. To the
best knowledge of the authors there are no analytic methods to
check this condition and we are only able to check $\bf H3$
numerically. To do that we  plot the real and imaginary part of
the the function of the variable $s$, $\theta(
\boldsymbol{U}s+\boldsymbol{K})$, where $\boldsymbol{U}$ is the
Ercolani-Sinha vector associated to the tetrahedron, i.e. $n=1$,
$m=0$. The plot shown in Fig. 2 confirms the validity of the
condition $\bf H3$ for this curve.

\begin{center}
\begin{figure}
\includegraphics[scale=0.4,angle=0]{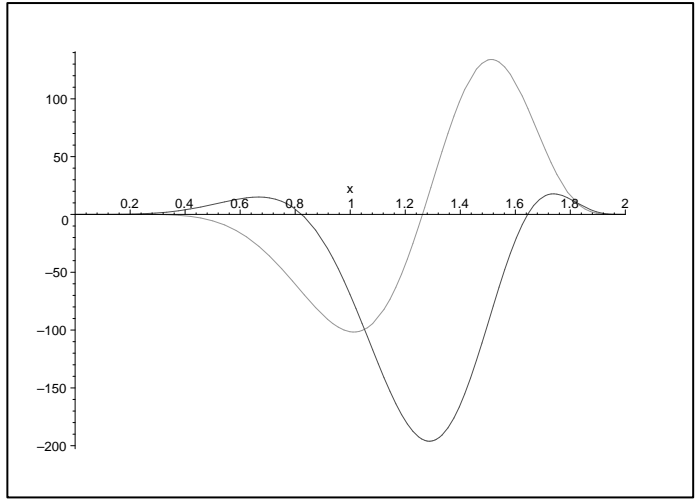}
\label{Fig.2}\caption{ Plot of the real and imaginary parts of the function
$\theta(\boldsymbol{U}s+\boldsymbol{K})$ for the case $n=1$, $m=0$}
\end{figure}
\end{center}

\vskip0.3cm {\bf 8. Uniqueness of the tetrahedral monopole.} Using
Ramanujan's hypergeometric relations many other solutions of
(\ref{Ramanujan}) were found \cite{bbg95} and from  each of these
one may construct curves satisfying the constraints $\bf H1$ and
$\bf H2$. Despite numerous attempts to find values for the
Ercolani-Sinha vectors different from the tetrahedrally symmetric
case just described no new solutions satisfying $\bf H3$ have been
found. We have conjectured that the solution corresponding to
arbitrary $n$, $m$ has $2(|n|-1)$ unwanted zeros in the interval
$s\in(0,2)$. For example, in the case $n=4$, $m=-1$, the plot of
$| \theta(\boldsymbol{U}s+\boldsymbol{K}) |$ is given in  Fig.3
which shows $6$ unwanted zeros. Therefore the corresponding value
of the parameter $b$ does not lead to a monopole curve. Although
unable to prove this general conjecture we are able to prove the
following theorem.

\begin{center}\begin{figure}
\includegraphics[angle=0,scale=0.4,trim =40 3 0 0 ]{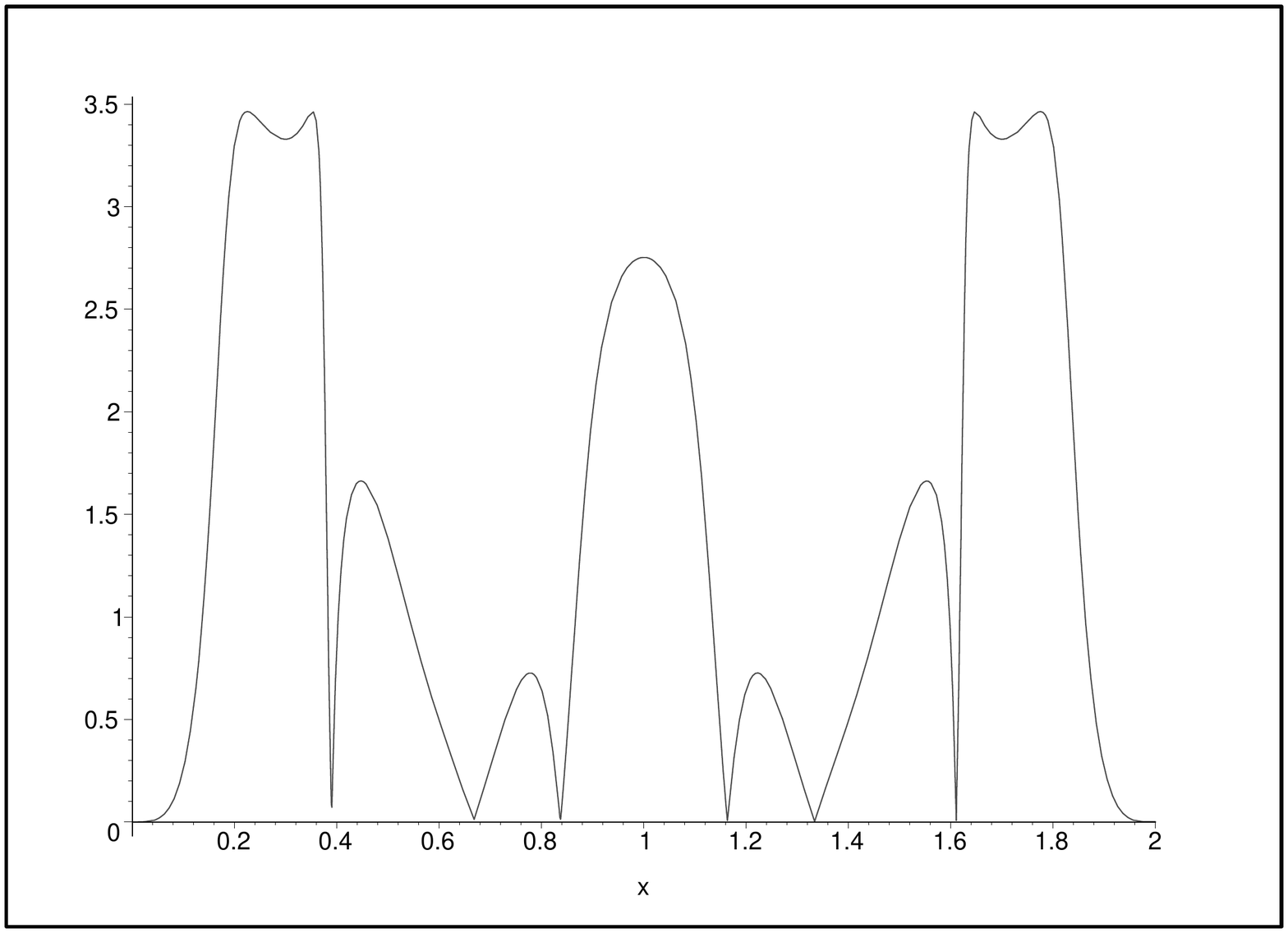}
\label{Fig.3}\caption{ Plot of the absolute value
$|\theta(\boldsymbol{U}x+\boldsymbol{K})|$ for the case $n=4$,
$m=-1$.}
\end{figure}
\end{center}
\begin{theorem}[\bf The uniqueness of the tetrahedral monopole \cite{bren10}]
The class of the monopole curves (\ref{bcurve})
consists of only two representatives,
\begin{equation}b=\pm 5\sqrt{2}, \qquad \chi=-\frac16\frac{\Gamma(1/6)\Gamma(1/3)}{2^{1/6}\pi^{1/2}}.
\label{terahedron}\end{equation} In other words there are no
monopoles associated to the curve (\ref{bcurve}) beyond those with
tetrahedral symmetry.
\end{theorem}

This is a statement for all integers $(m,n)\in \mathbb{Z}^2$.
Clearly this cannot be proven by resort to plots.   We will
demonstrate below the problem can be reduced to the analysis of
certain one-dimensional subsets in the plane. We are able to do
this by implementing one of the most remarkable achievements of
the theory of $\theta$-functions, namely the Schottky-Jung
proportionalities \cite{fay73}.

\vskip0.3cm {\bf 8.1 Schottky-Jung proportionality.} Schottky-Jung
theory permits the reduction of $\theta$-functions to
$\theta$-functions of lower genera for certain subspaces of the
Jacobi variety when a curve admits coverings. We consider here the
case of an unramified cover which may be associated to our family
of curves (\ref{C3}). Indeed our genus 4 curve
$\widehat{\mathcal{C}}$ covers 3-sheetedly a genus 2 curve
$\mathcal{C}$: $ \pi: \widehat{\mathcal{C}}\rightarrow \mathcal{C}
$ with
\begin{align} \widehat{\mathcal{C}}:&\;\;\eta^3+\chi(\zeta^6+b\zeta^3-1)=0, \\
 \mathcal{C}:& \;\; \nu^2=(\mu^3+b)^2+4, \end{align}
and $\nu=\zeta^3+1/\zeta^3, \mu =-\eta/\zeta$.
The Riemann-Hurwitz formula,
\[  2-2\widehat{g}=B+N(2-g) \]
shows that the cover $ \pi$ is unramified, i.e. $B=0$. More
generally one can associate to a class of $C_n$ symmetric curves
an $n$-sheeted unbranched cover of a hyperelliptic curve of genus
$n-1$ that is the spectral curve for the $su(n)$ affine Toda
theory \cite{b10}.

According to the Schottky-Jung theory (we are following here
\cite{fay73}) in the case of an unramified cover there exists a
basis in the homology group
  \[H_1(\hat{\mathcal{C}},\mathbb{Z})\ni (\mathfrak{a}_1,\ldots,\mathfrak{a}_4;\mathfrak{b}_1,\ldots,\mathfrak{b}_4) \]
admitting the automorphism $\sigma$ such that,
\begin{align}
\begin{split}
&\sigma\circ\mathfrak{a}_k
=\mathfrak{a}_{k+1},\quad\sigma\circ\mathfrak{b}_k=\mathfrak{b}_{k+1},\quad k=1,2,3\\
&\sigma\circ \mathfrak{b}_0=\mathfrak{b}_0,\qquad\sigma\circ \mathfrak{a}_0\sim\mathfrak{a}_0,
\end{split} \label{sigma}
\end{align}
(where $\sim$ means `homologous to'). The period matrices  of the
curves $\widehat{\mathcal{C}}$ and $\mathcal{C}$  are related by
\begin{align}
\hat{\tau}=\left( \begin{array}{cccc} a&b&b&b\\ b&c&d&d\\ b&d&c&d\\b&d&d&c \end{array} \right),\qquad
\tau =\left(  \begin{array}{cc}  \frac13a&b\\ b& c+2d  \end{array} \right). \label{FAperiods}
\end{align}

Remarkably under these conditions the following $\theta$-function
factorization occurs
\begin{theorem}[\bf The Fay-Accola theorem in the case of $g=4$]
In the case of the genus $g=4$  3-sheeted unramified covering of the genus two
curve the theta-factorization has the form
\begin{equation} \frac{ \theta(3z_1,z_2,z_2,z_2;\widehat{\tau})}
{
\theta(z_1,z_2;\tau)\theta(z_1+1/3,z_2;\tau)\theta(z_1-1/3,z_2;\tau)
}=\kappa, \label{FAproportionality} \end{equation} where
$\widehat{\tau}$ and $\tau$ are given in (\ref{FAperiods}) and
$\kappa$ is independent of $\boldsymbol{z}$.
\end{theorem}
We remark that the Fay-Accola theorem depends strongly on the
pull-back formula $(z_1,z_2)\rightarrow (3z_1,z_2,z_2,z_2)$,  but
the $\boldsymbol{z}$-argument of the genus four $\theta$-function
has the necessary form.

\begin{center}
\begin{figure}
\begin{minipage}{160pt}%\vspace*{-2mm}\hspace*{-7mm}
 \includegraphics[width=190pt]{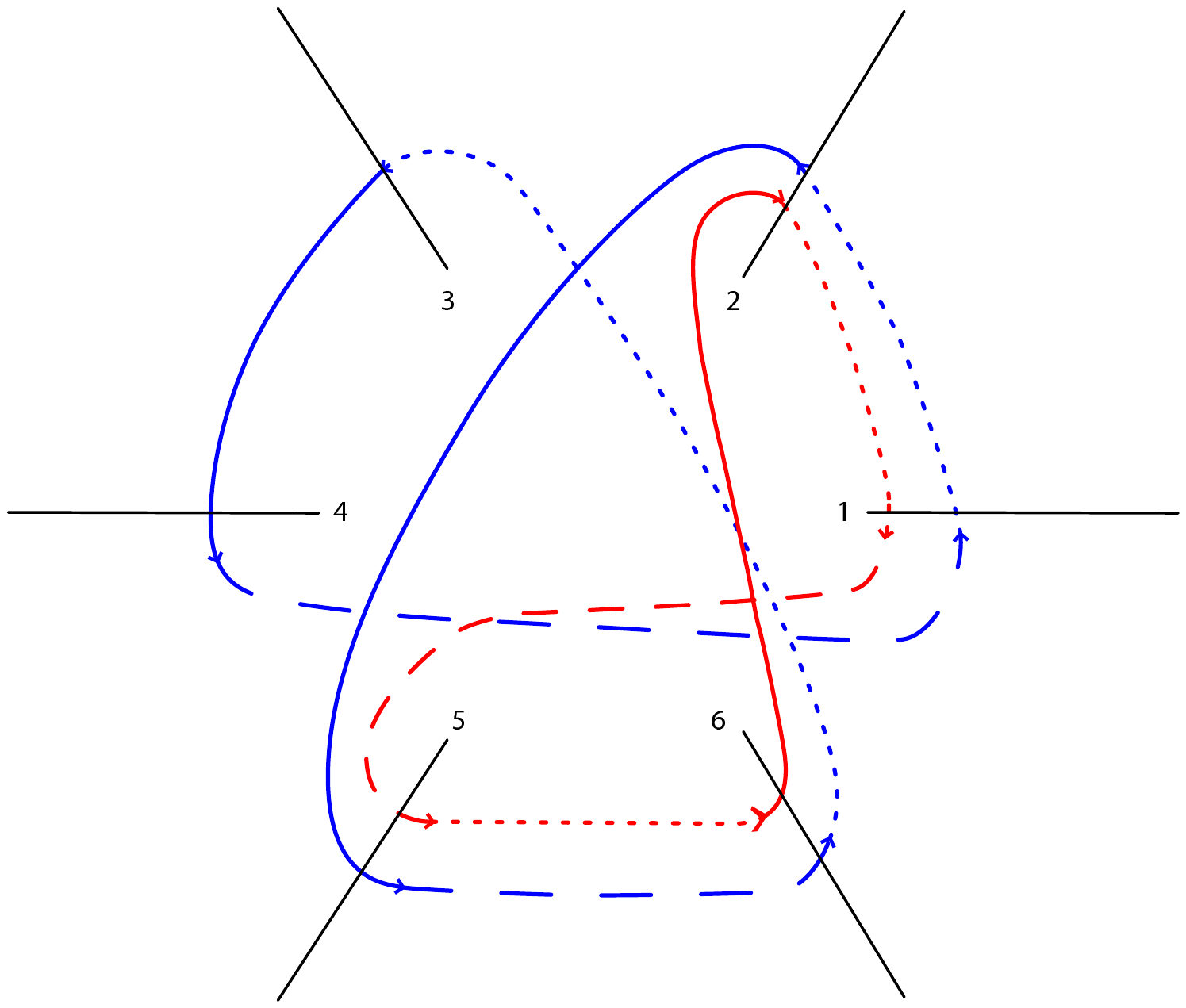}
\end{minipage}
\begin{minipage}{160pt}\vspace*{-0.4in}\hspace*{0.1in}
 \includegraphics[width=190pt]{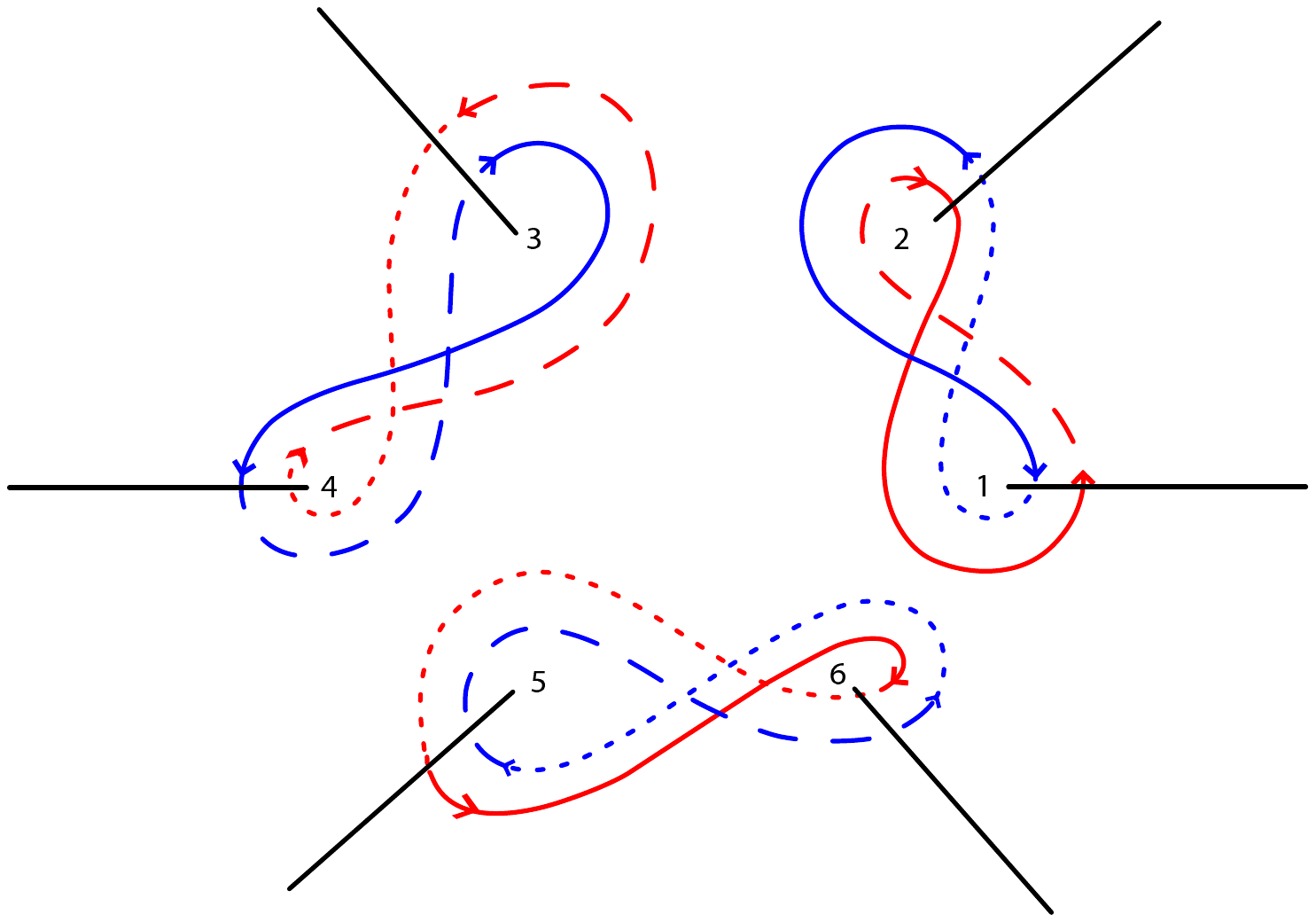}
\end{minipage}
\label{Fig.4}
\vspace*{-0.7in}
\caption{Symmetric homology basis of the curve (\ref{bcurve})}\end{figure}\end{center}

\vskip0.3cm {\bf 8.2 Homology transformation.} To implement the
Fay-Accola theorem we should first find a `cyclic' homology basis
(\ref{sigma}) for the curve (\ref{C3curve}). Such a basis is given
in Fig.4. Further, when the curve (\ref{C3curve}) is reduced to
(\ref{bcurve}) we wish to know  the symplectic transformation
between the homology basis given on the Fig.1 and that of Fig.4.
This will permit us to compare results obtained for this basis
with the previous ones of \cite{bren06}.

The cyclic homology basis was found in \cite{dav10,bde10} using
software developed by Northover \cite{nor10,nor10a}. This software
also provides us with the desired symplectic transformation
\cite{bn9}. Passing in our formulae for $\theta$-functions to the
cyclic homology basis we are able to reduce the analysis of the
vanishing of the genus four $\theta$-function to the analysis of
the vanishing of three genus two $\theta$-functions with arguments
shifted by $\pm\frac13$, as is evident in
(\ref{FAproportionality}).

\vskip0.3cm {\bf 8.3 Humbert variety.} Each of the aforementioned
genus two $\theta$-functions admits a further reduction to
elliptic $\theta$-functions. This is because the period matrix
$\tau$ matrix belongs to the so called {\em Humbert variety} that
is defined as follows. The Humbert variety $\mathcal{H}_{\Delta}$
consists of those period matrices $\tau$ of a genus two curve
$\mathcal{C}$ in the Siegel upper half-space that satisfy
\begin{align}\begin{split}
&q_1+q_2\tau_{11}+q_3\tau_{12}+q_4\tau_{22}
+q_5(\tau_{12}^2-\tau_{11}\tau_{22})=0 ; \\
 &q_i\in \mathbb{Z},\quad \Delta=q_3^2-4(q_1q_5+q_2q_4).
 \end{split} \label{humbert}
   \end{align}
It is known that in the case $\Delta=h^2$, $h\in \mathbb{N}$,
there exists a symplectic transformation $\mathfrak{S}$ that
reduces the period matrix to the quasi-diagonal form
\[ \mathfrak{S}:\tau\rightarrow \mathfrak{S}\circ\tau = \left( \begin{array}{cc} T_1&\frac{1}{h}\\
\frac{1}{h}&T_2   \end{array}  \right), \quad h\in \mathbb{N}.\]
The integer $h$ is the degree of the cover $\mathcal{C}$ over two
elliptic curves $\mathcal{E}$, $\mathcal{E}'$
\begin{align*}  \mathcal{E}'\leftarrow \mathcal{C}\rightarrow \mathcal{E}.
                                     \end{align*}
In the case we are considering the associated genus two curve is a
two-sheeted cover over an elliptic curve, i.e. $h=2$. The
underlying genus two curve has $D_6$ symmetry and is given by
Bolza's classification of genus two curves with many automorphisms
\cite{bol87}. Thus the genus two period matrix appearing in our
study is equivalent to
\begin{equation}
\left( \begin{array}{cc}  T& \frac12\\
\frac12& -\frac{1}{12T}       \end{array}  \right) \label{D6}
\end{equation}
where the complex variable $T$ is expressible in terms of the
Ercolani-Sinha vector and roots of unity. Thus a complete
reduction of the initial genus four $\theta$-function to elliptic
$\theta$'s occurs.

\begin{proposition} [\bf On the H3 condition  \cite{bren10}]  The vanishing
of the genus four $\theta$-function
\[ \theta(\boldsymbol{U}s+\boldsymbol{K};\widehat{\tau})=0\quad \text{for}\quad  s\in (0,2) \]
of the curve $\widehat{\mathcal{C}}$ given in (\ref{bcurve}) and
satisfying $\bf H1$ and $\bf H2$ occurs if and only if one of the
following three equalities is satisfied
\begin{align}\label{3conditions}
\frac{\vartheta_3}{\vartheta_2}\left( y\sqrt{-3}+\varepsilon \frac{T}{3}\vert T
\right)+(-1)^{\varepsilon}\frac{\vartheta_2}{\vartheta_3} \left( y+\varepsilon \frac{1}{3}\vert\frac{ T}{3} \right)=0,
\end{align}
where $\varepsilon=0,\pm 1$, and $$ y=\frac13 s(n+m), \quad T =
\frac{2\sqrt{-3}(n+m)}{2n-m}.$$
\end{proposition}

Therefore the function $  y=y(T)$ implicitly defined by
(\ref{3conditions}) provides the answer to the question of whether
or not $\bf H3$ satisfied. In studying this equation we found
(new?) $\theta$-constant relations
\[\frac{\vartheta_3}{\vartheta_2}\left(\frac{T}{3}\vert T\right)
=\frac{\vartheta_2}{\vartheta_3}\left(\frac13\vert\frac{T}{3}\right)\]
and
\[
\vartheta_4^2(0|T)\imath\sqrt{3}\frac{ \vartheta_1\left(\frac{T}{3}|T\right)
\vartheta_4\left(\frac{T}{3}|T\right) }{\vartheta_2^2\left(\frac{T}{3}|T\right)}+
\vartheta_4^2(0|\frac{T}{3})\imath\sqrt{3}\frac{ \vartheta_1\left(\frac{1}{3}|\frac{T}{3}\right)
\vartheta_4\left(\frac{1}{3}|\frac{T}{3}\right) }{\vartheta_3^2\left(\frac{1}{3}|\frac{T}{3}\right)}=0.
\]
Both relations can be proven by using Ramanujan's parametrization
of the Jacobian moduli of elliptic curves whose periods are $T$
and $T/3$, see \cite{lawd89} and \cite{bbg95}. The above
$\theta$-constant relations are used to analyze the plot in Fig.5.
They show that only in the two cases, when $(n+m)/(2n-m)=2$ and
$(n+m)/(2n-m)=1/2$ corresponding to the factors of $2$ and $1/2$
in the Ramanujan's hypergeometric relation (\ref{Ramanujan}), does
the $\theta$-divisor only intersect the boundaries of the segment
$[0,2]$ and no interior points. Therefore we can conclude that no
charge 3 monopoles exist for this class of curves beyond  those
with tetrahedral symmetry.

\begin{center}
\begin{figure}
\includegraphics[scale=0.4,angle=0]{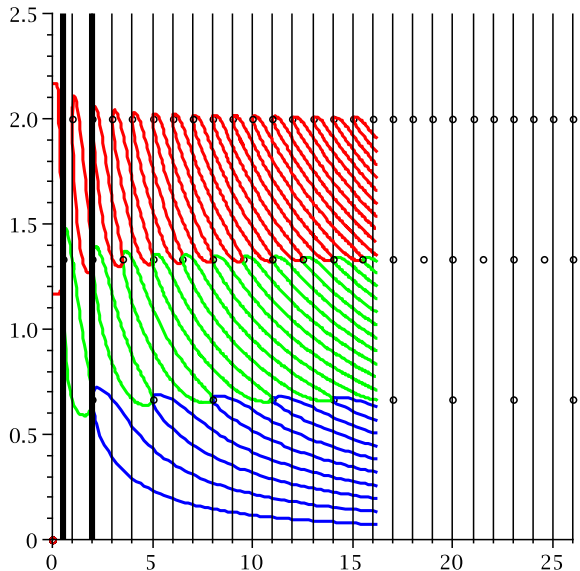}
\label{Fig.5} \caption{Three branches of the function $y(T)$ where $y/\rho$ is plotted against
$(n+m)/(2n-m)$. Here $T=2\sqrt{-3}(n+m)/(2n-m)$.}
\end{figure}
\end{center}

\vskip0.3cm {\bf 9. A new monopole curve.} Being armed with the
tetrahedral solution we are able to extend this result to the
general charge three monopole curve with $C_3$ symmetry given by
formula (\ref{C3curve}). First, by rescaling of the variables to
$(a,g):=(\alpha/\beta\sp{2/3},\gamma/\beta)$, one may recast the
Ercolani-Sinha constraints  to finding the $(a,g)$ such
\begin{equation}\label{agmes}
0=\oint\limits_{\boldsymbol{\mathfrak{c}}}\frac{d\,
X}{Y},\qquad Y^2=(X^3+a\, X+g)^2+4,
\end{equation} for a cycle
$\boldsymbol{\mathfrak{c}}$ specified by the solutions $n=1$,
$m=0$ and $n=1$, $m=1$. The remaining Ercolani-Sinha constraint
simply determines $\beta$ in terms of $(a,g)$. Thus starting with
the points $(a,g)=(0,5\sqrt{2})$ and $(a,g)=(0,-5\sqrt{2})$ we can
find the line in the real $(a,g)$-plane along which $\bf H1$, $\bf
H2$, $\bf H3$ are satisfied.

That is one of the results of A.D'Avanzo \cite{dav10} and
\cite{bde10}. The outline of the method is as follows. The
integral is evaluated using the genus two arithmetic-geometric
mean (AGM) which generalizes the Gaussian AGM method for
calculating complete elliptic integrals of the first kind. The
genus two AGM method as presented in the Bost-Mestre article
\cite{bostmestre88} deals mainly with real branch points and a
modification of this method to the case of a real curve and
complex branch points was developed. Using this AGM we may quickly
determine those $a$ and $g$ for which $(\ref{agmes})$ is
satisfied.

The corresponding plot is given in Fig.6. This curve reproduces
the asymptotic behavior predicted by \cite{hmm95}. One can see
that the cusp point $(3,0)$ appears on the plot. The curve here
reduces to the rational curve
\begin{equation}  y^2=(x^2+4)(x^2+1)^2,  \label{degenerate}  \end{equation}
i.e. becomes singular. We remark that this behavior of the
solution curve is consistent with Sutcliffe's prediction that the
curve (\ref{degenerate}) describes a configuration of three
unit-charge monopoles with dihedral $\mathrm{D}_3$ symmetry,
constituting an asymptotic state for a 3-monopole configuration
(cf. eq.(4.16) in \cite{Sut97}).

\begin{figure*}
\begin{center}
\includegraphics[height=250pt]{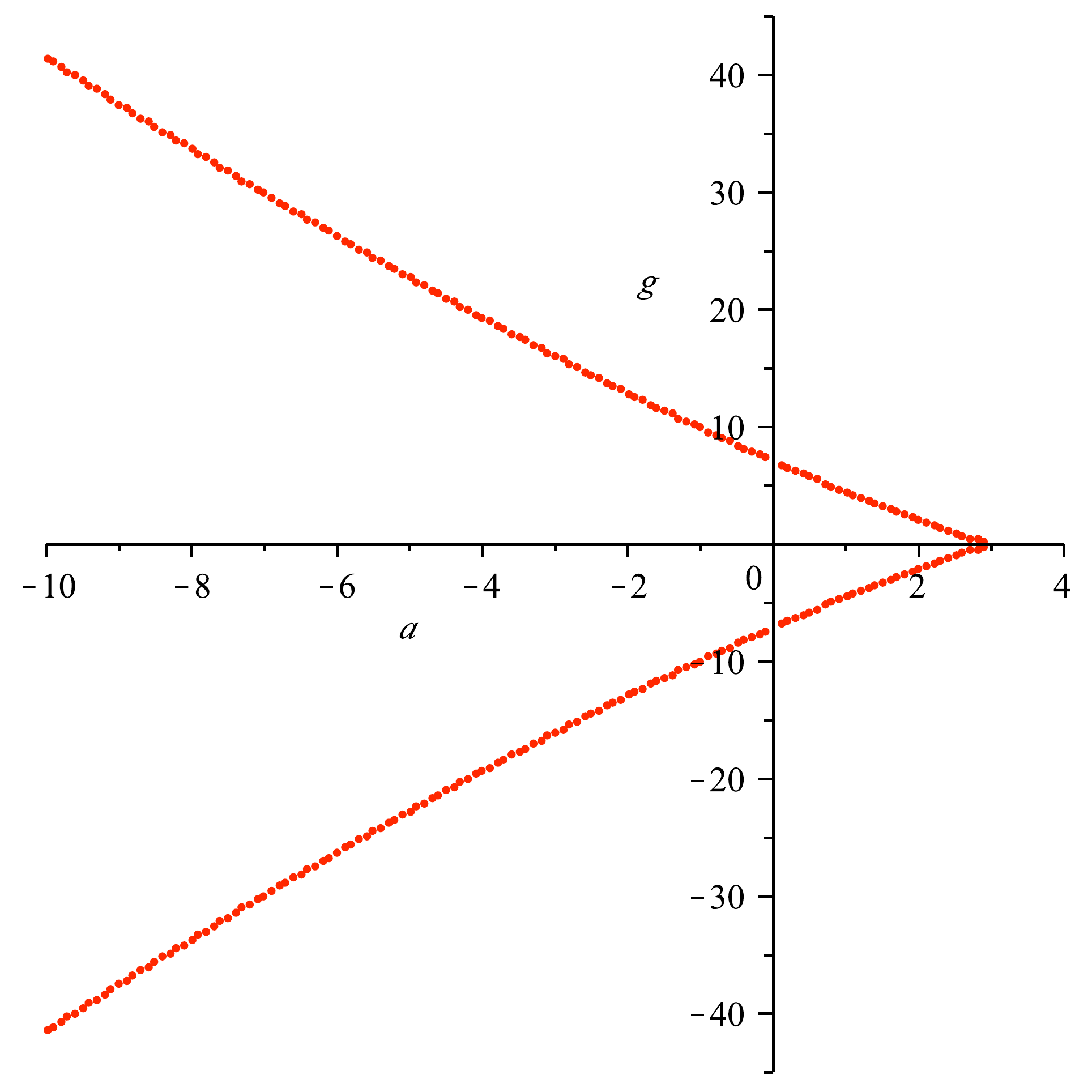}
%\end{minipage}
\caption{Solutions to the Ercolani-Sinha
constraints}\label{ESsoln}
\end{center}
\end{figure*}

\vskip0.3cm {\bf 10. Discussion.} In this note we have
concentrated on the finding the algebraic curve that satisfies
Hitchin's constraints for the spectral curve of monopole. We
succeeded in finding a new one-parameter family of charge 3
monopole curves of the form
$\eta^3+\alpha\eta\zeta^2+\beta\zeta^6+\gamma\zeta^3-\beta=0$ that
admits $C_3$ symmetry.

But this is only the first step of the construction. The ultimate
aim is to find the Higgs field $\Phi$ and gauge fields $a_i$ in
closed analytic form. To the best knowledge of the authors, with
the exception of the $n=1$ and $n=2$ axially symmetric cases, no
such analytic expressions have yet been found. Our program is to
calculate such analytic quantities and the work presented here is
part of that program. Knowledge of the monopole curve allows one
to develop the algebro-geometric integration of the Nahm equation
as was done in \cite{ersi89}. These last results were improved in
\cite{bren06}. A common perception is that the ADHMN construction
requires the numerical solution of the Weyl equations with
potentials provided by the Nahm data. We are seeking the
$\theta$-functional integration of the Weyl equations themselves.
In \cite{bren09} we use an ansatz of {Nahm} \cite{nahm82} which
reduces the integration of the $2n$-th order system of ODE of the
Weyl equations to an $n$-th order ODE system that is equivalent to
the linear spectral problem of the Lax representation for Nahm
equation. Therefore the algebro-geometric solution to the Weyl
equation is given in terms of a Baker-Akhiezer function of the
Nahm equation whose spectral parameter is a function of monopole
coordinates. We believe that such a program is realizable for
higher charge monopoles and, in particular, for the one-parameter
family of trigonal curves reported here. This approach in the case
of a non-axially symmetric  charge $2$ monopole is now the focus
of our attention and the results will be published elsewhere.

\vskip0.3cm {\bf Acknowledgments.} This note is written on the
basis of talks delivered to the XXIX Workshop on Geometric Methods
in Physics, Bialowieza - 27.06-03.07.2010. One of the authors, VZE
is grateful to the HWK, Hanse-Wissenschaftskolleg (Institute of
Advanced Study) in Delmenhorst for the grant in January-July 2010
when his talk was prepared.

\providecommand{\bysame}{\leavevmode\hbox
to3em{\hrulefill}\thinspace}
\providecommand{\MR}{\relax\ifhmode\unskip\space\fi MR }
\providecommand{\MRhref}[2]{%
  \href{http://www.ams.org/mathscinet-getitem?mr=#1}{#2}
}
\providecommand{\href}[2]{#2}

\end{document}